\documentstyle[12pt]{article}
\begin{document}
\baselineskip=16pt
\newcommand{\newc}{\newcommand}
\newc{\be}{\begin{equation}}
\newc{\ee}{\end{equation}}
\newc{\bea}{\begin{eqnarray}}
\newc{\eea}{\end{eqnarray}}
\newc{\bean}{\begin{eqnarray*}}
\newc{\eean}{\end{eqnarray*}}
\newc{\eqn}[1]{(\ref{#1})}
\newc{\gsim}{\lower.7ex\hbox{$\;\stackrel{\textstyle>}{\sim}\;$}}
\newc{\lsim}{\lower.7ex\hbox{$\;\stackrel{\textstyle<}{\sim}\;$}}
\def\stopr{\widetilde t_R}
\def\slr{\widetilde l_R}
\def\hinot{\widetilde H^0_2}
\def\hinob{\widetilde H^0_1}
\def\bino{\widetilde B^0}
\def\wino{\widetilde W^0_3}
\def\gluino{\widetilde g}
\def\mgluino{m_{\gluino}}
\def\msl{m_{\widetilde l}}
\def\msq{m_{\widetilde q}}
\newc{\mz}{m_Z}
\newc{\mw}{m_W}
\newc{\abund}{\Omega h_0^2}
\newc{\mchi}{m_\chi}
\newc{\mcharone}{m_{\charone}}	\newc{\charone}{\chi_1^\pm}
\newc{\mhalf}{m_{1/2}}
\newc{\mzero}{m_0}
\newc{\muzero}{\mu_0}
\newc{\mtop}{m_t}
\newc{\mbot}{m_b}
\newc{\mgut}{M_X}
\newc{\ie}{{\it i.e.}}
\newc{\etal}{{\it et al.}}
\newc{\eg}{{\it e.g.}}
\newc{\etc}{{\it etc.}}

\def\thefiglist#1{\section*{Figure Captions\markboth
{FIGURE CAPTIONS} {FIGURE CAPTIONS}}\list
{Figure \arabic{enumi}.}
{\settowidth\labelwidth{Figure #1.}\leftmargin\labelwidth
\advance\leftmargin\labelsep
\usecounter{enumi}}
\def\newblock{\hskip .11em plus .33em minus -.07em}
\sloppy}
\let\endthefiglist=\endlist
\def\NPB#1#2#3{Nucl. Phys. B {\bf#1} (19#2) #3}
\def\PLB#1#2#3{Phys. Lett. B {\bf#1} (19#2) #3}
\def\PLBold#1#2#3{Phys. Lett. {\bf#1B} (19#2) #3}
\def\PRD#1#2#3{Phys. Rev. D {\bf#1} (19#2) #3}
\def\PRL#1#2#3{Phys. Rev. Lett. {\bf#1} (19#2) #3}
\def\PRT#1#2#3{Phys. Rep. {\bf#1} C (19#2) #3}
\def\ARAA#1#2#3{Ann. Rev. Astron. Astrophys. {\bf#1} (19#2) #3}
\def\ARNP#1#2#3{Ann. Rev. Nucl. Part. Sci. {\bf#1} (19#2) #3}
\def\MODA#1#2#3{Mod. Phys. Lett. A {\bf#1} (19#2) #3}

\begin{titlepage}
\begin{flushright} RAL-93-003\\UM-TH-33/92\\ December 1992\\
hep-ph/9301267\\
\end{flushright}
\vskip 2cm
\begin{center}
{\bf\large Implications for Minimal Supersymmetry from
Grand}	
\vskip 4pt
{\bf\large
Unification and the
Neutralino Relic Abundance}
\vskip 1cm
{\bf  R. G. Roberts\\}
\vskip 2pt
{\it
Rutherford Appleton Laboratory,\\
Chilton, Didcot, Oxon,
OX11 OQX, England}\\
\vskip .25cm
{\it and}
\vskip .25cm
{\bf Leszek Roszkowski\\}
\vskip 2pt
{\it Randall Physics Laboratory, University of Michigan,\\ Ann Arbor,
MI 48190, USA\\
leszek@leszek.physics.lsa.umich.edu}\\
\end{center}
\vskip .5cm
\begin{abstract}
We examine various predictions of the minimal supersymmetric standard
model coupled to minimal supergravity.  The
model is characterised by a small set of parameters at the
unification scale.
The supersymmetric particle spectrum at low energy and the
spontaneous
breaking of the standard model itself are then generated radiatively.
The previously considered
predictions of the model now include the neutralino relic density
which in turn
provides bounds on the scale parameters. We find a remarkable
consistency among several different constraints which imply all
supersymmetric particle masses preferably within the reach of future
supercolliders (LHC and SSC). The requirement that the neutralino be
the dominant component
of (dark) matter in the flat Universe provides a {\em lower} bound
on the spectrum of supersymmetric particles
beyond the reach of LEP, and most likely
also the Tevatron and LEP 200.

\end{abstract}
\end{titlepage}
\setcounter{page}{2}
\section{Introduction}

The minimal extension of the standard model~\cite{susyrev} which
corresponds to a
softly-broken supersymmetric $SU(3)\times SU(2)_L \times U(1)_Y$ at
the scale
$\mgut$ where the gauge couplings unify (as recently confirmed
by LEP~\cite{amaldi}) provides a very attractive
and
economic description of physics beyond the standard
model.  It is
possible to
specify a small number of parameters at the unification scale and the
low
energy effective theory is then determined simply by the radiative
corrections.
In particular the spontaneous breaking of electroweak symmetry is
radiately
generated due to the presence of supersymmetry soft-breaking terms
through the mass squared of one of the two Higgs doublets being
driven negative at the scale $Q\simeq{\cal O}(\mz)$ by the Yukawa top
quark
coupling~\cite{{gsymbreak}}.
In terms of the starting parameters at the GUT scale a detailed
spectrum of
the supersymmetric (SUSY) states is completely determined.
Even in the simplest SUSY scenario one meets considerable uncertainty
related to the presence of both the superheavy states around the GUT
scale and, more importantly, new supersymmetric states above
$\mz$~\cite{bh,meshkov,rr}.
Clearly the corresponding threshold corrections around $\mgut$ depend
on which unified
group or superstring scenario the minimal SUSY model is embedded
into. This inherent uncertainty would weaken the predictive
power
of the theory and, as in a previous paper~\cite{rr}, we perform a
minimal analysis where such corrections are ignored but corrections
from supersymmetric states above $\mz$ are treated with particular
care~\cite{rr}. Similarly,
the important constraint coming from the limits on the proton
decay~\cite{protondecay}
depends on
the choice of a specific GUT model and will not be discussed
here.

In this letter, we extend the previous analysis~\cite{rr} to include
the
predictions for the relic abundance of the lightest neutralino $\chi$
which
is typically the lightest supersymmetric particle (LSP) of the model.
The neutralino LSP has
long been identified~\cite{ehnos} as one of the leading candidates
for dark matter
in the Universe~\cite{dmrev,kt}. It is neutral, weakly interacting,
stable (if $R$-parity is valid) particle and its relic density is
typically consistent with present cosmological expectations.
We examine the predictions for $\chi$  from the minimal
supersymmetric standard model (MSSM) and compute the annihilation
cross
sections which requires the detailed knowledge of the whole SUSY
spectrum.
Consequently we can relate values of the neutralino relic abundance
to values
of the parameters $\mhalf, \mzero, \muzero$ --- the common gaugino
mass, the
common scalar mass and the higgsino mass at $\mgut$.
The lower limit on the age of the Universe provides an upper bound on
the
relic abundance of matter, and in particular of dark matter which is
believed to be a dominant mass component of the Universe.
We
can therefore use the dark matter abundance constraint to
derive bounds on the ranges of $\mhalf, \mzero, \muzero$ and in turn
get
constraints on the masses of all the SUSY particles.

In fact we can combine the dark matter constraint with others which
are
either
phenomenological (values of $\alpha_s (\mz), \mtop, \mbot)$ or
theoretical
(avoiding mass hierarchy problem) and examine the consistency of
trying to
satisfy several of these constraints simultaneously.  We conclude
that one can
indeed achieve such consistency quite naturally.
More interestingly, we find that this happens for
the ranges of the fundamental
parameters $\mhalf$, $\mzero$, $\muzero$, and thus also
masses of the supersymmetric particles, all preferably
within the few hundred GeV mass range and thus
{\em well within the reach} of the SSC and the LHC but typically
{\em above the reach}  of LEP, the Tevatron, and LEP 200. The lower
limit on supersymmetric particle masses comes from the dark matter
constraint as will be discussed in section 3.

The LSP for which we find sufficiently large values of the relic
abundance
to explain at least DM in the galactic halos ($\abund\gsim 0.025$)
invariably comes out to be almost
gaugino-like (bino-like) consistent with the
conclusions of some previous analyses~\cite{chiasdm,sugradm,dn}. It
was first noticed in
Ref.~\cite{chiasdm}
that a higgsino-like LSP is somewhat disfavoured as it corresponds to
a
high scale of supersymmetry breaking, typically exceeding 1 TeV, and
thus a gaugino-like LSP was selected as a unique candidate for DM.
More
recently, it has been
shown~\cite{dn,japan}
that for the higgsino-like neutralinos additional effects
(co-annihilation with the next-to-lightest
neutralino and the lightest chargino, see sect.~3) have a dramatic
effect of reducing the LSP relic
abundance below any interesting level. Here we find that
higgsino-like
LSPs are also largely excluded by the
current lower bound on the mass of the top quark.

Overall, the LSP relic abundance
constraint, combined with
the other constraints narrows down the allowed ranges of $\mhalf,
\mzero,
\muzero$ considerably. We find that the region $\mhalf\gg\mzero$ is
excluded by the lower bound on the top mass,
while in the region $\mhalf\ll\mzero$ the LSP relic abundance is too
large ($\abund>1$). Furthermore, the requirement that the LSP provide
enough
missing mass in the flat ($\Omega=1$) Universe can be fulfilled only
in
a relatively narrow band of comparable values of $\mhalf$ and
$\mzero$ and for $1<\muzero/\mzero\lsim {\rm a~few}$.

In the next section we briefly review and update the procedure
used in deriving the low-energy spectrum from a limited number of
basic parameters at the GUT scale. In section 3 we calculate the
neutralino relic density and compare it with other constraints on
the parameter space. We conclude with final remarks in section~4.

\section{Solutions of the MSSM}

We consider the MSSM in the context of a unified theory.  At
the
compactification scale $\mgut$ where the three couplings of
$SU(3),SU(2),U(1)$
have a common value $\alpha_X$ the SUSY parameter space is
characterised by
the common values of the gaugino masses $\mhalf$, the common value of
the soft mass terms of the squarks, sleptons and Higgs bosons
$\mzero$,
and by $\muzero$, the mass parameter of the Higgs/higgsino bilinear
term
in the superpotential. (The suffix 0
denotes values at $\mgut$.) In addition there are two parameters
characterising the soft terms proportional to the superpotential
terms: $B_0$ in the bilinear term $B_0\muzero$, and a common
trilinear soft parameter $A_0$ which multiplies the Yukawa terms.
Also one should
include at least the Yukawa couplings $h_{t0}, h_{b0}, h_{\tau 0}$ at
$\mgut$ to consider as parameters in principle.

However we can reduce this apparently
unmanageable host of parameters
down to a manageable set as follows.
Firstly the coefficients $A_0$ and $B_0$ are set to zero,
in the spirit of string-derived versions of the model~\cite{rr}.
Below the
scale $\mgut$,
$B$ grows to a finite positive value and generally reaches a maximum
and
may even decrease to negative values.  The values of $\mgut$ and
$\alpha_X$
are determined by the unification of the gauge couplings.  Their
precise values
for each solution are computed by an iterative procedure discussed in
Ref.~\cite{rr} since the running of the gauge couplings depends on
knowing the individual
SUSY thresholds which in turn depend on all the parameters including
$\mgut$
and $\alpha_X$ themselves.  This procedure requires the measured
values of
$\alpha_1 (\mz), \alpha_2 (\mz)$ from LEP but the value of $\alpha_3
(\mz)$ must be adjusted to achieve the required unification for each
solution.
Another adjustment is to choose $h_{t0}$ such that the running Higgs
mass
squared $m^2_2(Q)$ takes on the precise value (negative and
${\cal O}(\mz^2$)) at
$Q=\mz$ needed to give the required spontaneous breaking of
electroweak
symmetry, \ie,
\be
(m^2_1-m^2_2) + (m^2_1+m^2_2) \cos 2\beta = - \mz^2 \cos 2\beta
\label{gbcondition}
\ee
where $m_1,m_2$ are the running masses of the Higgs doublets
coupling to
down- and up-type quarks respectively.  Here the ratio of the Higgs
{\it v.e.v.}s
$v_2/v_1 = tan\beta = \cot\theta$ with $\beta$ related to $\mu$ and
$B$ by
$\sin 2\beta =2B\mu /(m^2_1+m^2_2)$.  The running of $m^2_2$ and
therefore the satisfying of eq.~(\ref{gbcondition}) is controlled by
the value of $h_{t0}$.
Actually the other significant Yukawa couplings $h_b, h_\tau$ should
be included in this running
of $m^2_2$ but in order to achieve eq.~(\ref{gbcondition}) in a
controllable way we drop them
which is justified as long as $\tan\beta$ is not too large
($\tan\beta\ll\mtop/\mbot$).

Thus each solution is specified by the values of the three parameters
$\mhalf$, $\mzero$, $\muzero$.  Each solution then provides at low
energies
specific
values for the quantities $\alpha_s (\mz)$, $\tan\beta,\mu$, gaugino
masses $M_1,M_2,M_3$, squark masses, slepton masses, Higgs masses,
Higgsino masses and top quark mass $\mtop =(\sqrt{2}h_t
\mw/g)\sin\beta$.
Relaxing the constraints $A_0=0, B_0=0$ affects the resulting value
of
$\tan\beta$ mostly --- see the analysis of Ref.~\cite{rr}, and so
quantities which
depend sensitively on $\tan\beta$ at low energies are, in principle,
less
precise, in our procedure.  From Ref.~\cite{rr} we see that, in
general, $\tan\beta >2$ even when $A_{0}, B_0$ are allowed to vary
within values of ${\cal O}(m_0$).

Another quantity associated with each solution is the ratio
$\mbot/m_\tau$,
assuming that this ratio is unity at $\mgut$.  Thus we include also
$h_{b0}=h_{\tau 0}$
as another parameter in the running of the Yukawa couplings, and
obtain
a specific value for $\mbot$ for each solution.  Apart from the above
phenomenological constraints on the solutions we have the standard
constraints that the Higgs potential be bounded, \ie,
\be
|\sin 2\beta|<1,\;\;
m^2_1m^2_2<\mu^2 B^2
\ee
and that all the physical mass squared be
positive.

The strongest constraint for insisting that the SUSY spectrum is
relatively
light comes from the `naturalness' argument~\cite{rr,bg} which
regards
the need to tune the value
of $h_{t0}$ to a very high precision in order for $m^2_2$ to take the
exact
value given by eq.~(\ref{gbcondition}) at the scale $Q=\mz$.  A
measure
of this `fine
tuning'
problem is the fine tuning constant $c$ defined by~\cite{rr}
\be
c = \frac{\delta \mw^2}{\mw^2}/\frac{\delta h^2_t}{h^2_t}
\label{cdef}
\ee
so that absence of fine tuning means $c \approx 1$. Approximately
we have~\cite{rr}
\be
c \approx \frac{m_0^2 + \mu_0^2 + k \mhalf^2}{m_Z^2}
\label{finetuning}
\ee
A reasonable limit to the degree
of precision needed would be $c \lsim {\cal O} (10)$ and
consequently the
typical
SUSY mass cannot be many times greater than $\mz$.

We illustrate the various constraints by showing the values of
$\mtop$ and $\mbot$ in Fig.~\ref{figone}a and
$\alpha_s(\mz)$  and $c$ in Fig.~\ref{figone}b as a function of
$\mhalf$ and $\mzero$ for a fixed ratio $\muzero/\mzero=2$.
The variation with $\muzero/\mzero$ will be discussed later.
The regions marked CDF and LEP are excluded by the CDF searches for
the top ($\mtop\gsim 91$ GeV) and the LEP searches for charginos
($\mcharone\gsim 46$ GeV), respectively.
We see from Fig.~\ref{figone}a  that the current `experimental' value
for $\mbot$ (in the $\overline{\rm MS}$ scheme),
$\mbot(2\mbot)= 4.25\pm 0.1$ GeV~\cite{bottom}, implies a rather
heavy
top
quark ($\mtop\gsim 150$ GeV) for the values of the input
parameters
$\mhalf,\mzero$ and $\muzero$ roughly within the 1 TeV limit. On the
other hand, beyond that range the resulting value of $\mbot$ is
consistent with $\mtop\lsim 150$ GeV. Larger values of the input
mass
parameters are, however, disfavoured by the fine-tuning constraint
and
the current bounds on $\alpha_s=0.122\pm0.010$ (based on analysis of
jets
at LEP)~\cite{altar}
as we can see from Fig.~\ref{figone}b.
We also note that the uncertainty on $\tan \beta$ arising from
allowing
$A_0$ and $B_0$ to be non-zero (discussed above) would imply that
$\mtop$
could be smaller by a further 10\%.

One
can see immediately that demanding $c\lsim {\cal O} (10)$ forces
one
to consider
only values of $\mhalf,\mzero$ up to a few hundred GeV.  This was the
conclusion of the previous analysis~\cite{rr}.  Thus unification of
the couplings
demands a value of $\alpha_s(\mz)$ close to the values extracted from
jet
analyses at LEP.

To summarise so far, the solutions obtained for the MSSM with the
inclusion of
electroweak symmetry allow a fairly restricted region of the
parameters
$\mhalf,\mzero,\muzero$ which is consistent with all the above
constraints, \ie,
$\mhalf,\mzero\lsim$ 200 GeV, $\muzero\lsim$ 400 GeV.
We will comment on the restrictions on the ratio $\muzero/\mzero$
later.

\section{The Neutralino Relic Abundance}

The knowledge of the whole mass spectrum of both the ordinary and
supersymmetric
particles allows one to reliably compute the relic abundance of
the lightest supersymmetric particle (LSP)
as a candidate for the dark matter in the Universe.

At the outset we note that, in the parameter space not already
excluded by LEP and CDF, we find that it is the lightest of the four
neutralinos that is {\em always} the LSP.
Another potential candidate for the LSP, the sneutrino, has been now
constrained by LEP to be heavier than about 42 GeV
and, if it were the LSP, its contribution to the relic abundance
would
be of the order of $10^{-4}$, and thus uninterestingly small. In the
analysis presented here, the sneutrino is typically significantly
heavier than the lightest neutralino. Typically, it is not even the
lightest sfermion.

The actual procedure of calculating the relic abundance has been
adequately described in the literature and will not repeated here.
We use the technique developed
in Ref.~\cite{swo} which allows for a reliable (except near poles and
thresholds) computation of the thermally
averaged annihilation cross section in the non-relativistic limit and
integrating the Boltzmann equation.

In the early Universe the LSP pair-annihilated into ordinary matter
with
total mass not exceeding $2\mchi$. In calculating the LSP relic
density
one needs to include all possible final states. Lighter $\chi$s
annihilate only
(except for rare radiative processes) into
pairs of ordinary fermions via the exchange of the $Z$ and the Higgs
bosons, and the respective sfermions. As $\mchi$ grows new final
states
open up: pairs of Higgs bosons, gauge and Higgs bosons, $ZZ$ and
$WW$,
and $t\bar t$. We include all of them in our analysis.

Generally one considers $\abund>1$ as incompatible with the
assessed lower bound of about 10 Gyrs on the age of the Universe or,
in
more popular terms, as corresponding to too much mass in the
Universe~\cite{kt}. Many astrophysicists strongly favour the value
$\Omega=1$ (or very close to one), corresponding to the flat
Universe, either because of cosmic inflation
or for aesthetical reasons. Moreover, there is growing evidence that,
on a global scale, the mass density indeed approaches the critical
density, as well as that most of the matter in the Universe is
non-shining and non-baryonic~\cite{kt}.
If one assumes that the LSP is the dominant component of dark matter
in the flat ($\Omega=1$) Universe then one typically expects
\be
0.25\lsim\abund\lsim0.5,
\label{flatuniv}
\ee
where the biggest uncertainty lies in our lack of knowledge of the
Hubble parameter $h_0$ to better than a factor of two. As we will
see shortly, varying somewhat the bounds in
eq.~(\ref{flatuniv}) will not significantly alter our conclusions.

We present in Fig.~\ref{figone}c the relic abundance of the LSP
and compare it with the other results shown before in
Figs.~\ref{figone}a and~\ref{figone}b. Several features can
be immediately noticed.

Firstly, most of the region corresponding to larger values of
$\mzero$ (roughly $\mzero\gsim$ 500 GeV)
is cosmologically excluded
as it corresponds to $\abund>1$. The relic abundance generally
decreases with decreasing $\mzero$ reaching very low values of
$\abund$ (0.025, or less) for $\mzero$ roughly below 200 GeV,
especially for $\mhalf>\mzero$.
It is worth noting that the region favoured by cosmology,
eq.~(\ref{flatuniv}),
takes a shape of a relatively narrow band running roughly parallel to
the border of the area excluded  by $\abund>1$. The contour
$\abund=0.1$ shows how quickly $\abund$ decreases with decreasing
$\mzero$ but also limits from below the region where the LSP relic
abundance is reasonably large.

It is interesting to see what mass and compositions of the LSP
correspond to
its relic abundance favoured by cosmology. We remind the reader that,
in minimal supersymmetry, the lightest neutralino
and its three heavier partners $\chi^0_i$ ($i=1,...,4$)
are the physical (mass) superpositions of
higgsinos $\hinob$ and $\hinot$, the fermionic partners of the
neutral Higgs bosons, and of two gauginos $\bino$ and $\wino$,
the fermionic partners of the neutral gauge bosons
\be
\chi\equiv
\chi_1^0=N_{11}\widetilde W^3+N_{12}\widetilde B+N_{13}\widetilde
 H_1^0+N_{14}\widetilde H_2^0.
\label{chilsp}
\ee
In distinguishing the gaugino-like and higgsino-like regions it
is convenient to use the gaugino purity $p=N_{11}^2+N_{12}^2$. In
particular, the LSP is almost a pure
bino where $p_{bino}\equiv N_{12}^2$
is close to one.
In Fig.~\ref{figone}d we show the bino purity of the LSP. (The
gaugino
purity is almost identical.) Remarkably, we find that the band
favoured by cosmology corresponds to the LSP being almost a pure bino
($\sim95\%$) up to very large values of $\mhalf$. We also find that
that higgsino-like LSPs are incidentally almost entirely excluded by
the lower
bound on the top quark of 91 GeV. (The contour of equal gaugino and
higgsino contributions almost coincides with the contour $\mtop=91$
GeV.) It was also noticed in Ref.~\cite{dn} that for a heavy top
constraints from radiative gauge symmetry breaking exclude
higgsino-like LSPs.
(With the expectation for $\mtop$ to be actually much heavier than 91
GeV a larger cosmologically uninteresting region is likely to be
ruled out.)
The LSP mass contours are almost vertical in the gaugino region with
$\mchi$ growing with $\mhalf$, and
almost horizontal in the higgsino region with $\mchi$ increasing with
$\mzero$. Again, the lines meet in the narrow sub-diagonal region
where
the LSP is both a  gaugino and a higgsino.

Since higgsino-like LSPs in our analysis not only give very little DM
but also are practically excluded by the CDF top searches, we need
not
worry about the additional effect of the higgssino-like LSP
`co-annihilation'~\cite{gs} with the next-to-lightest neutralino and
the lightest chargino which has been recently shown to significantly
reduce the LSP relic density~\cite{dn,japan}.
We have explicitly verified that {\it all} solutions for
which
co-annihilation of the LSP with $\chi_2$ and $\chi^\pm_1$ is
important
lie in the region excluded by $\mtop\geq 91$~GeV. Thus neglecting the
effects of co-annihilation is justified.

The LSP relic abundance in the allowed region is mostly dominated
by its annihilation into
fermionic final states, although in a few cases the Higgs final
states
contributed comparably. We thus do not expect that the radiative
corrections to the Higgs masses due to the heavy top~\cite{radcorrs}
would noticeably modify our results~\cite{sugradm}. We also found
that the lightest sfermion is
either $\stopr$ or $\slr$, in agreement with Ref.~\cite{dn}, except
in the (mostly excluded by LEP) region of small $\mzero$ and $\mhalf$
where it is the sneutrino.

We now pass to combine the band favoured by cosmology with
the mass contours of the top and the bottom quarks. This is shown
in Fig.~\ref{figtwo}. We see that the region where the LSP gives
the dominant contribution to the matter density of the flat Universe
(marked $\Omega=1$) crosses  the estimated value of
the bottom quark mass ($\mbot=4.25\pm0.1$ GeV) for $\mtop$ broadly
between 160 GeV and 180 GeV. Remarkably, this happens for
150 GeV$\lsim\mhalf,\, \mzero\lsim$ 400 GeV, the range also strongly
favoured
by constraints from $\alpha_s$ and fine tuning.

When the ratio
$\muzero/\mzero$ is decreased, the relic abundance contours
generally move towards larger values of $\mzero$ as do the contours
for $\mtop$ and $\mbot$. For $\muzero/\mzero=1$
the favoured range of the bottom quark mass of about 4.25 GeV lies
entirely within the cosmologically excluded region $\abund>1$.
It also becomes harder to reconcile this region with the fine tuning
constraint and with a value of $\alpha_s(\mz)$ close to 0.122. On
the other hand as $\muzero/\mzero$ increases, $\mbot = 4.25$ takes us
to a region of larger $\mhalf$ and lower $\mzero$ while the contours
relic abundance remain relatively unchanged.
The area consistent with the constraints of $\mbot$, $\mtop$
correspond to lower values of the relic abundance, $\abund \lsim
0.25$.
If we increase $\muzero/
\mzero$ still further the region of overlap for the constraints of
$\mbot$, $\mtop$, and $\Omega=1$ shrinks and leads us the region of
larger fine tuning and smaller $\alpha_s$.
We thus conclude that the combination of all the above constraints
selects the range $1\lsim\muzero/\mzero\lsim {\rm a~few}$.

In the selected range all the Higgs bosons, squarks and sleptons, as
well as the
gluino, are significantly lighter than 1 TeV and thus are bound to
be found at the LHC and SSC.

However, the expectation that the LSP dominates the dark matter
relic density is a natural one. (In minimal supersymmetry no other
particle can even
significantly contribute to the missing mass.) It then implies
a significant {\em lower} bound on the spectrum of supersymmetric
particle masses. We see from Figs.~\ref{figone}d and ~\ref{figtwo}
that the LSP masses favoured by all the constraints lie in the range
\be
60 ~{\rm GeV}\lsim\mchi\lsim 200 ~{\rm GeV},
\label{lsprange}
\ee the upper limit being also expected
in the
minimal supersymmetric model~\cite{chiasdm,bg} on the basis
of naturalness. Similarly, we find
\bea
150 ~{\rm GeV}\lsim\mcharone\lsim300~{\rm GeV}\\
200 ~{\rm GeV}\lsim\msl\lsim500 ~{\rm GeV}\\
250 ~{\rm GeV}\lsim\msq\lsim850 ~{\rm GeV}\\
350 ~{\rm GeV}\lsim\mgluino\lsim900 ~{\rm GeV}.
\label{massranges}
\eea
The heavy Higgs bosons are roughly in the mass range between 250 GeV
and 700 GeV. Of course, lower values of all these masses correspond
to less fine tuning and larger values of $\alpha_s$.
The lightest Higgs boson tree-level mass invariably
comes out close to $\mz$; its one-loop-corrected
value~\cite{radcorrs} is then roughly in the range 120 to 150 GeV.
By comparing Figs.~\ref{figone}b and ~\ref{figtwo}
we also find  $0.116\lsim\alpha_s(\mz)\lsim 0.120$. (Larger values
of $\alpha_s$ are also disfavoured by considering threshold
corrections
at the GUT scale~\cite{meshkov}.)

Thus, if the LSP is indeed the dominant component of DM in the flat
Universe, supersymmetric particles are probably beyond the reach
not only of LEP but also the Tevatron and LEP
200~\cite{chiatlep2,sugradm,dn,texas}.
We note, on the other hand, that smaller ranges of supersymmetric
particles are not firmly excluded but would correspond to the LSP
contributing only a fraction of the critical density. We also note
that we do not claim to have done a fully exhaustive search of the
whole
parameter space. In fact, Drees and Nojiri~\cite{dn} have found in
certain
extreme cases (rather large values of $A_0$) squarks even somewhat
lighter that 200 GeV and a lower limit $\mzero>40$ GeV. We find that
the condition $\Omega=1$ requires in our case significantly larger
values of
$\mzero$ ($\mzero\gsim$ 150 GeV), in agreement with
Refs.~\cite{sugradm,texas}. However, we do not consider it to be in
contradiction with the mentioned results of Ref.~\cite{dn} but a
reflection of somewhat different assumptions at the GUT scale and
methods of deriving the
supersymmetric mass spectra.
We do not expect that the procedure adopted here would produce
substantially modified results by performing a finer search of the
parameter space.

\section{Conclusions}

Our basic conclusions for the
neutralino relic abundance and the associated implications for
the supersymmetric mass spectra are generally consistent with several
other recent analyses.
We do find that cosmologically attractive LSP is almost purely
bino-like ($\sim95\%$) and lies in the range
60 GeV$\lsim\mchi\lsim$ 200 GeV. Moreover, as first noted in
Ref.~\cite{chiatlep2} and confirmed in
Refs.~\cite{sugradm,dn,texas},
if the LSP dominates the dark matter in
the (flat) Universe then the expected ranges of chargino, slepton
and Higgs boson masses lie
beyond the reach of LEP 200.
The associated ranges of gluino and squark
masses
then exceed the reach of the Tevatron but should be discovered at
the SSC and/or the LHC.

Generally, we find it very reassuring that, in the simplest and most
economic supersymmetric scenario, a careful analysis of the
implications
of several
different (and independent) constraints, including
the DM constraint, which result from
the grand unification conditions, leads to
a supersymmetric spectrum accessible to the next generation of
accelerators.

\section*{Acknowledgments}

We thank Graham Ross and Gordon Kane for inspiration and numerous
discussions. This work was supported in part by the US Department
of Energy.

\newpage

\newpage
\section*{Figure Captions}
\vskip 0.1cm

\begin{figure}[htb]
\vspace{0.1cm}
\caption{In the plane ($\mhalf,\mzero$) for the fixed ratio
$\muzero/\mzero=2$ we show: in window a) the mass contours of the top
and the bottom quarks (solid and short-dashed lines, respectively);
in window b) the contours of $\alpha_s(\mz)$ (solid) and the measure
$c$
of fine-tuning (dots), as discussed in the text; in window c) the
relic
abundance $\abund$ of the LSP; and in window d) the mass contours of
the LSP (solid) and the lightest chargino (dashed) at 50, 100, 150,
200, 500, and 1000 GeV, starting from left,
and the contribution (dots)
of the
bino to the LSP composition (bino purity, as discussed in the text).
In all the windows thick
solid lines delineate regions experimentally excluded by the CDF
(marked CDF) where $\mtop<91$ GeV and by the LEP experiments (LEP)
where the lightest chargino is lighter than 46 GeV. In window c) we
also mark by $\abund>1$ the region cosmologically excluded (too young
Universe). The thin band between the thick dashed lines in window c)
corresponds to the flat Universe ($\Omega=1$), as discussed in the
text. In window d) the region excluded by CDF almost coincides with
the bino purity of 50$\%$ or less.
}
\label{figone}
\end{figure}

\begin{figure}[htb]
\vspace{0.1cm}
\caption{We show a blow-up of the down-left portion of the plane
($\mhalf,\mzero$) from the previous figure for the same fixed ratio
$\muzero/\mzero=2$.  We combine the mass contours of the top and the
bottom quarks with the ones  of the LSP relic mass density. We use
the
same textures as in Fig.~1 but we also show
(two medium-thick short-dashed lines) the contours $\mbot=4.15$ GeV
and
4.35 GeV which reflect the currently favoured range of the mass of
the
bottom quark (see text). We see that they cross the cosmologically
favored region (thick long-dashed lines) marked $\Omega=1$ at roughly
150 GeV$\lsim\mhalf,\, \mzero\lsim$ 400 GeV and for $\mtop$ broadly
between 150 GeV and 180 GeV.
}
\label{figtwo}
\end{figure}
\end{document}